\documentclass[12pt,a4paper]{article}

\usepackage{url}
\usepackage{amsmath}
\usepackage{amssymb}
\usepackage{graphics}
\usepackage{graphicx}
\usepackage{xcolor}
\usepackage[symbol]{footmisc}

\usepackage{kmath,kerkis} % The order of the packages matters; kmath changes the default text font\\

\begin{document}

		\begin{center}
		
		{{{\Large\bf Dynamics and Statistics of Weak Chaos in a 4--D  Symplectic Map}} }\\
		
		\vspace{0.5cm}
		{\large\sf Tassos Bountis\footnote[1]{E-mail address: tassosbountis@gmail.com}$^{, 1}$, Konstantinos Kaloudis$^{2}$, and Helen Christodoulidi$^{3}$}

		\centerline{$^{1}$ Center for Integrable Systems, P.G. Demidov Yaroslavl State University, 150003, Yaroslavl, Russia}
		\centerline{$^{2}$ Department of Statistics and Actuarial-Financial Mathematics, University of the Aegean,}
		\centerline{Samos, 83200, Greece}
		\centerline{$^{3}$ Department of Mathematics, Lincoln University, Lincoln, LN6 7TS UK}
		\vspace{0.2cm}
	\end{center}

\begin{abstract}
The important phenomenon of ``stickiness'' of chaotic orbits in low dimensional dynamical systems has been investigated for several decades, in view of its applications to various areas of physics, such as classical and statistical mechanics, celestial mechanics and accelerator dynamics. Most of the work to date has focused on two-degree of freedom Hamiltonian models often represented by two-dimensional (2D) area preserving maps. In this paper, we extend earlier results using a 4--dimensional extension of the 2D McMillan map, and show that a symplectic model of two coupled McMillan maps also exhibits stickiness phenomena in limited regions of phase space. To this end, we employ probability distributions in the sense of the Central Limit Theorem to demonstrate that, as in the 2D case, sticky regions near the origin are also characterized by ``weak'' chaos and Tsallis entropy, in sharp contrast to the ``strong'' chaos that extends over much wider domains and is described by Boltzmann Gibbs statistics. Remarkably, similar stickiness phenomena have been observed in higher dimensional Hamiltonian systems around unstable simple periodic orbits at various values of the total energy of the system.  
\end{abstract}
\section{Introduction}

The behavior of nonlinear dynamical systems described by differential and difference equations has been a topic of intense interest for several decades \cite{GuckHolm1983,Wiggins1990,Ott2002,LichtLieb2013,Strogatz2015}. As is   well-known, one the most important questions in this field concerns the distinction between solutions of the equations that are called ``regular'', since their evolution can be predicted for long times, and those termed ``chaotic'', whose time evolution becomes unpredictable after relatively short times. This is typically decided by calculating the Lyapunov exponents, measuring the distance between two nearby solutions, represented by trajectories (or orbits) in the $2N-$ dimensional phase space of the system \cite{Skokos2010}, with $N$ position and $N$ momentum variables, with time as the single independent variable. If none of the Lyapunov exponents is positive we call the orbit {\it regular}, while if at least one exponent is positive we call it {\it chaotic}.

But is this ``duality'' between order and chaos all there is? While there is no uncertainty about regular orbits,  it has been realized that ``chaos'' is a lot more subtle to describe by a simple definition. One possibility is to  study chaotic phase space domains from a statistical point of view, in terms of correlations and probability distributions. If these correlations decay exponentially away from a chaotic orbit, one might adopt a Boltzmann Gibbs (BG) thermodynamic description of the dynamics (as in the case of an ideal gas) and look for Gaussian probability functions (pdfs) to describe the associated statistics. What happens, however, if correlations decay by power laws and the pdfs of positions and/or momenta are no longer Gaussian? What would that imply about the corresponding chaotic behavior? 

One such widely known example occurs in cases of ``stickiness'', where chaotic orbits of generally low-dimensional dynamical systems tend to remain confined for very long times trapped within thin chaotic layers surrounding regions of regular motion \cite{LichtLieb2013,ContHar2010,KatPatCont2013,ContVoEfth1997,KovErd2009,BoMaAnt2012}.
Remarkably, this phenomenon does not occur only in low dimensions. It has also been observed in multidimensional Hamiltonian lattices \cite{BouSko2012,AntBaBou2010,AntBoBa2011,ChTsaBo2014,ChBoTsaDro2016,ChBoDro2018}, often in cases where chaotic regions arise around simple periodic orbits, when they have just turned unstable, as the total energy of the system is increased. 

Regarding dynamical systems in discrete time, it is well--known that 2D Poincar\'{e} maps describe intersections of the orbits of a 2-degree of freedom continuous dynamical system with a 2D surface of section \cite{LichtLieb2013}. Thus, one may consider directly area preserving transformations of a plane onto itself to study the qualitative features of such maps \cite{TirBor2016}. 

One famous model in this regard is the 2D McMillan (2DMM) area preserving, non-integrable map \cite{RuBoTsa2012}. It may be interpreted as describing the dynamics of focusing a ``flat'' proton beam in a circular particle accelerator model describing the repeated passage of a ``flat'' beam through a periodic sequence of thin nonlinear lenses \cite{TurchScan1991}:
\begin{eqnarray}\label{2DMMmap}
x_{n+1}&=&y_n  \nonumber \\  
y_{n+1}&=&-x_n+\frac{2K y_n}{1+y_n^2}+\mu y_n,
\end{eqnarray}
where $x_n$ and $y_n$ represent a particle's position and momentum at the nth crossing of a focusing element, while $\mu$, and $K$ are physically important parameters. Note that the Jacobian of the transformation is unity, so that (\ref{2DMMmap}) is area-preserving and thus may represent the conservative (Hamiltonian) dynamics of proton beams whose radiation effects are considered negligible \cite{TurchScan1991}. If $\mu=0$ the map is integrable, as it possesses a constant of the motion given by the one parameter family of curves \cite{HJN2016}:
$$x^2_n+ y_n^2+x^2_ny_n^2-2Kx_ny_n=const.$$

In \cite{RuBoTsa2012}, (\ref{2DMMmap}) was studied following a nonextensive statistical mechanics approach, based on the nonadditive Tsallis entropy $S_q$ \cite{Tsallis2010}. According to this approach, the pdfs optimizing $S_q$, under appropriate constraints, are $q$--Gaussian distributions that represent quasistationary states (QSS) of the dynamics, with $1<q<3$ ($q=1$ being the Gaussian). As was shown in \cite{RuBoTsa2012}, there are several cases of $K > 1$ and $\mu>0$ parameters, where the chaotic layer around a saddle point at the origin does {\it not} satisfy BG statistics associated with ``strong chaos'', but is well described by a $q>1$-Gaussian pdf, associated with ``weak chaos''. 

It is, therefore, natural to ask whether similar phenomena of spatially limited, weakly chaotic dynamics occur in 4D symplectic maps, such as one encounters e.g. in 3-degree-of-freedom hamiltonian systems commonly encountered in problems of celestial mechanics, see e.g. \cite{ContHar2010,KatPatCont2013,ContVoEfth1997,KovErd2009} and particle accelerator dynamics \cite{BouSko2006a,BouSko2006b}. 

In this paper, we extend for the first time the above approach to study 4D McMillan (4DMM) maps of the form
\begin{eqnarray} \label{4DMMmap}
x_{n+1}&=&-x_{n-1}+\frac{2K_1 x_n}{1+x_n^2}+\mu x_n-\epsilon x_ny_n^2 \nonumber \\  
y_{n+1}&=&-y_{n-1}+\frac{2K_2 y_n}{1+y_n^2}+\mu y_n-\epsilon x_n^2y_n  
\end{eqnarray}
where $x_n, y_n$ represent horizontal and vertical deflections of the proton beam as it passes through the $nth $ focusing element and study the chaotic domain arising about the origin of (\ref{4DMMmap}), using values of $K_1$, $K_2$ and $\mu$ for which the origin is unstable. Note that (\ref{4DMMmap}) is symplectic, as the evolution of $x_n$ and $y_n$ is determined by a potential function $V(x_n,y_n)$, whose partial derivatives with $x_n$ and $y_n$ respectively yield the two equations of (\ref{4DMMmap}).

We choose suitable $K_1$ and/or $K_2$ values, for fixed $\mu>0, \epsilon>0$ small, such that the origin is (linearly) unstable and calculate the pdfs of the rescaled sums of $N$ iterates of the map, in the sense of the Central Limit Theorem, in the large $N$ limit for large sets of initial conditions. We then relate our results to specific properties of the phase space dynamics of the maps and distinguish cases where the pdfs represent long--lived QSS described by $q$-Gaussians.

We begin by describing in Section \ref{Section2} the statistical methods used in this paper to obtain the pdfs describing our data in all cases of the 4DMM map studied here. Next, in Section \ref{Section3}, we apply this analysis to find weak chaos characterized by $q-$ Gaussian pdfs, for different parameter values connected with an unstable fixed point at the origin of our 4DMM map. We end with our conclusions in Section \ref{Section4}.

\section{Statistical analysis of weak chaos\label{Section2}}

Before turning to the 4DMM mapping studied here, we first carried out the same computations for the 2DMM map (\ref{2DMMmap}) and compared them to results depicted in Fig. 3(a) of \cite{RuBoTsa2012}. Employing the same choices of initial conditions and the same number of iterations, we verified that we obtain practically identical results. 

For the benefit of the reader, we state that the approach we follow here is to evaluate the solution $x_n,y_n$, $n=0, \ldots, N$ of the 4DMM map (\ref{4DMMmap}) and construct probability distributions for $x_n$ (similarly for $y_n$) of appropriately large rescaled sums  $S_{j}(N)$ obtained by adding the corresponding $N$ iterates
$$
S_{j}(N)=\sum_{n=0}^N x_n^{(j)}
$$
where $j$ refers to the $j$--th realisation, taking values from 1 to the total number of initial conditions $N_{ic}$. 
As in \cite{RuBoTsa2012}, we generate the centered and rescaled sums
\begin{equation}\label{Sums}
s_j(N) \equiv \frac{ S_j(N)- \mu_j(N) } { \sigma_N}=\left(\sum_{n=0}^N x_n^{(j)}-\frac{1}{N_{i c}} \sum_{j=1}^{N_{i c}} \sum_{n=0}^N x_n^{(j)}\right) / \sigma_N
\end{equation}
where $\mu_j(N)$ is the mean value and  $\sigma_N$ the standard deviation of $S_j(N)$ over $N$ iterations
$$
\sigma_N^2=\frac{1}{N_{i c}} \sum_{j=1}^{N_{i c}}\left(S_j(N)- \mu_j(N) \right)^2=\left\langle S_j^2(N)\right\rangle- \mu_j^2(N),
$$
where $< \cdot >$ denotes averaging over $N$ iterations. We thus find many cases, where the obtained empirical distributions are well--described by a $q$-Gaussian distribution of the form

\begin{equation}\label{qGaussdef}
P\left(s_j(N)\right)= \dfrac{\sqrt{\beta}}{C_q} \left[1+\beta(q-1) s_j^2(N)\right]^{1/{1-q}}
\end{equation}
where $q$ is regarded as an indicator measuring the divergence from the classical Gaussian distribution, $\beta$ is the `inverse temperature'  fitting parameter and $C_q$ is a normalizing constant.

To describe the statistical properties of the above rescaled sums of the system, we employ standard parameter estimation techniques. Specifically, we are interested in identifying the $q$-Gaussian distribution that best describes the observed data. One of the most widely used methods for such estimations, is the Maximum Likelihood Estimator (MLE) \cite{rossi}. This is a \textit{parametric} method typically used for statistical fitting among distributions belonging to the same family, e.g. the family of Gaussians parameterized by their mean and standard deviation or the family of $q$-Gaussians parameterized by $\left(q,\beta\right)$.

The main idea behind the MLE is that the most suitable distribution (of a given family) describing a given data set, is {\it the most probable} to describe the observed data. More formally, we are interested in maximizing the likelihood function, which describes ``how likely'' it is to observe a certain random sample, for the various values of the unknown parameters of the assumed statistical model.

To determine the likelihood function $p\left(\theta | \mathbf{X} \right)$, we first calculate the joint probability function of the observed sample $\mathbf{X} = \left(X_1=x_1,\ldots,X_n=x_n\right)$ as a function of the parameters of the problem, $\theta= \left(\beta,q\right) \in \mathbb{R}^{+} \times \left[1,3\right)$. Then, the MLE is the value of $\theta \in \Theta$ that maximizes the likelihood function, i.e. $\hat{\theta} = \arg\max_{\theta \in \Theta}p\left(\theta | \mathbf{X} \right) $. For computational purposes, it is convenient to maximize the logarithmic likelihood function, which for a $q$-Gaussian statistical model has the form:
\[
l_{\mathbf{X}}\left(\beta, q\right) = \sum_{i=1}^{n} \log \dfrac{\sqrt{\beta}}{C_q} \left[1+\beta(q-1)x_i^2\right]^{1/{1-q}}.
\]
In all simulations that follow, we perform our numerical optimization using the so-called ``nlm'' (nonlinear minimization) command of the $R$ software for statistical computing \cite{rlang}.

An alternative approach to derive optimal $q$-Gaussian parameters is to apply nonlinear least-square fitting to binned estimates of the probability density (via histograms), using such methods as Gauss-Newton (see e.g. \cite{white}). However, from a statistical point of view, it is more accurate to use MLE instead of curve-fitting estimates, as the MLE are theoretically guaranteed, under general (regularity) conditions, to have such desirable properties, as efficiency, consistency and asymptotic normality \cite{mle_ref}. For an interesting discussion of the comparison between curve-based estimates and MLEs for the case of q-Exponential distributions, we refer the reader to Shalizi \cite{shalizi1}.

\section{Evidence of weak chaos in 4DMM maps\label{Section3}}

\subsection{Weak chaos in an example of the 4DMM map}

\begin{figure}
	\centering
	\includegraphics[scale=0.65]{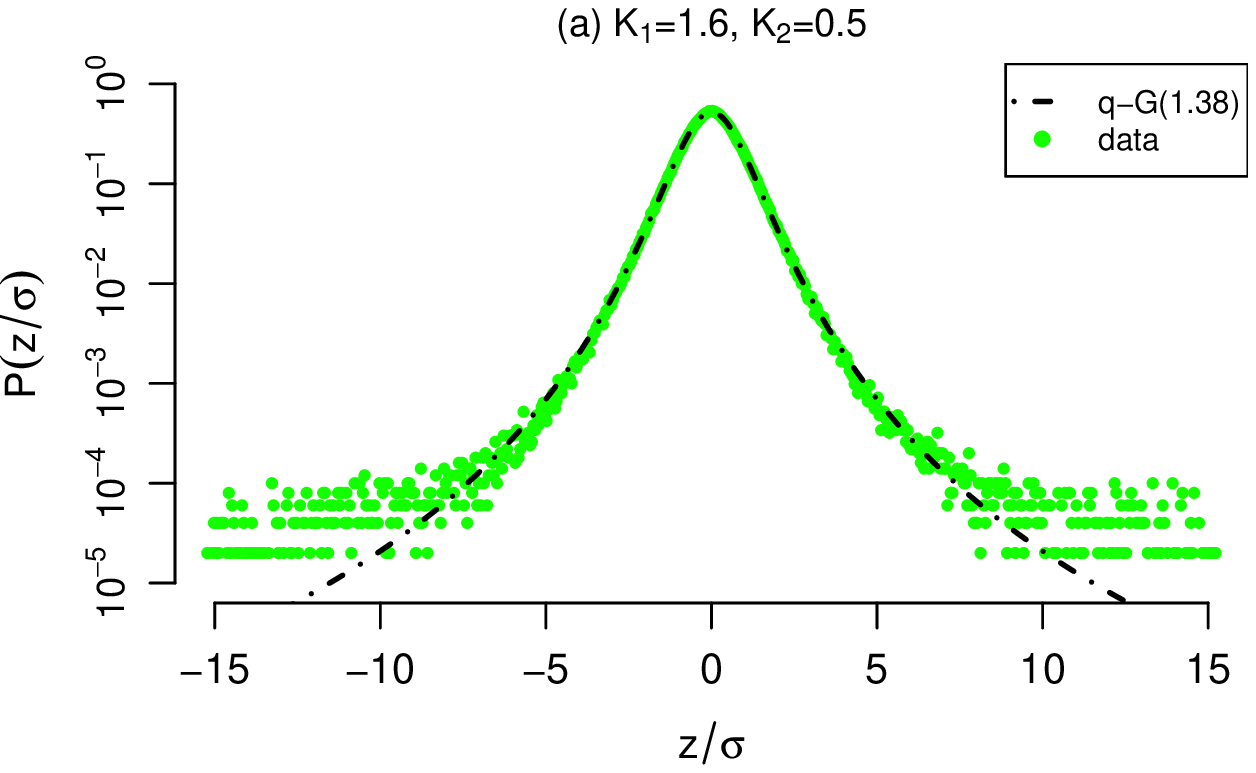}\\
	\includegraphics[height=0.2\textheight]{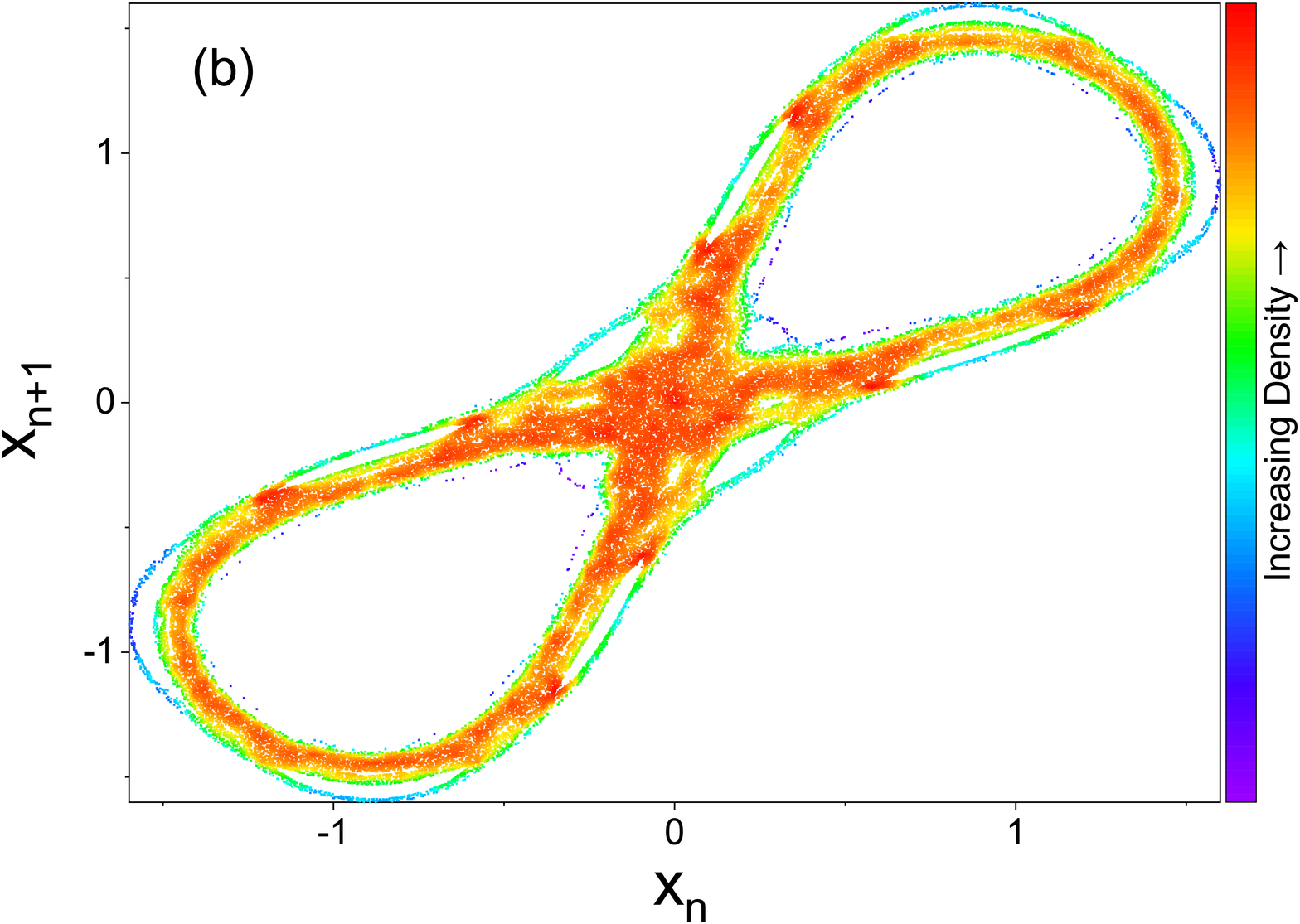}
 \includegraphics[height=0.2\textheight]{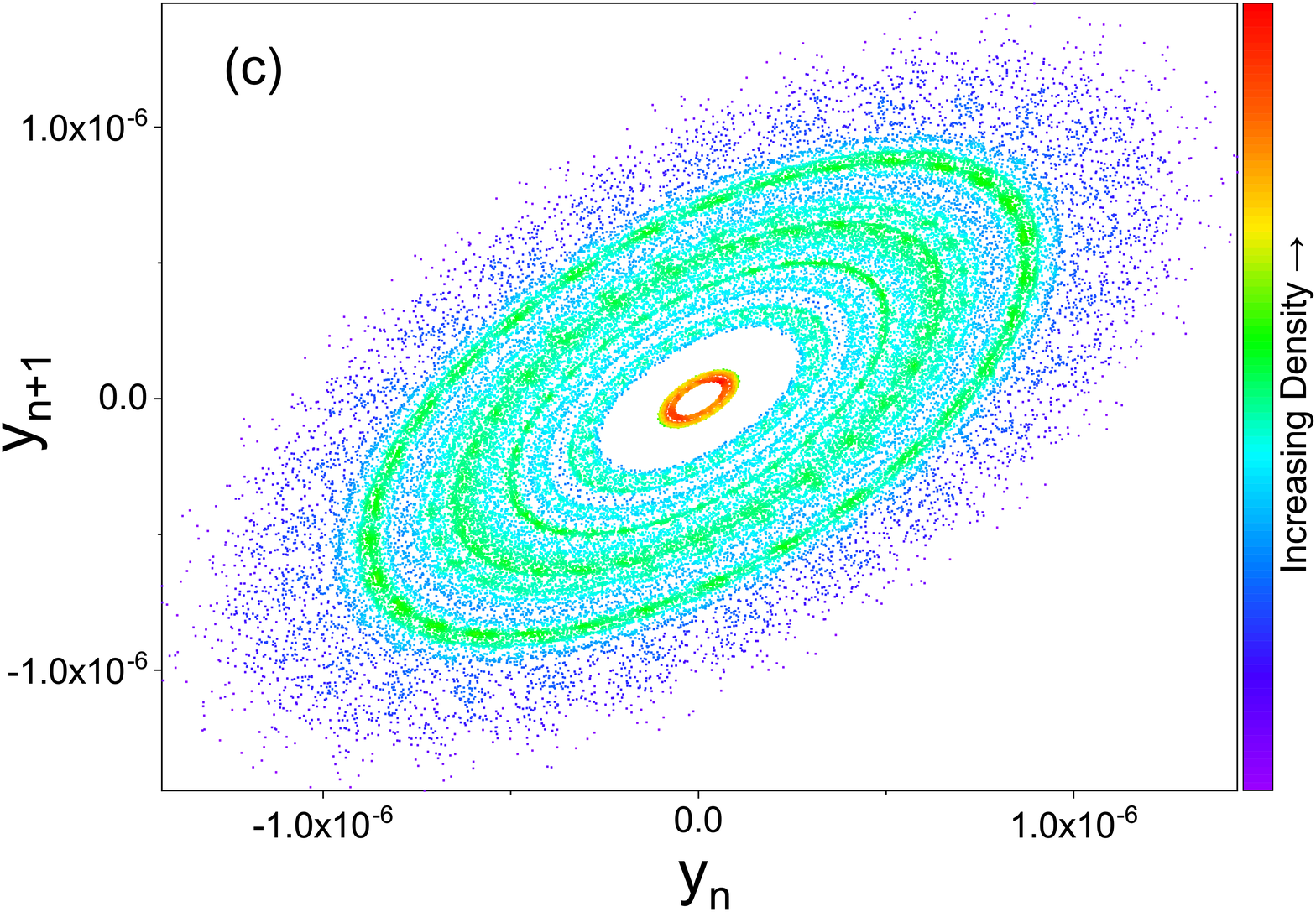}
	\caption{(a) The computation of the pdf for the $x_n$ variable in (\ref{4DMMmap}) with parameters $K_1=1.6, K_2=0.5$, $\mu=0.2$, and $\epsilon=0.01$. The dashed line represents an optimal fitting of the data by a $q$-Gaussian function (\ref{qGaussdef}) with $q=1.38$ and $\beta=1.19$. 
		(b) The 2D phase space plot of the $x_n, x_{n+1}$ projections of the 4D  map (\ref{4DMMmap}) for the orbits and parameters used in (a), while (c) shows the 2D phase space projection in the $y_n,y_{n+1}$ plane variables.\label{fig0}} 
\end{figure}

We start by fixing the values of $\mu=0.2$ and $\epsilon=0.01$, which we will use throughout the paper, as they do not significantly affect the results.  Observe now in our Fig. \ref{fig0}(a) a typical example of an optimal pdf of a $q$-Gaussian obtained for the choice of parameters $K_1=1.6,K_2=0.5$. This is a case we shall call hyperbolic--elliptic (HE), referring to the first 2D map in (\ref{4DMMmap}) having a hyperbolic fixed point at the origin, and the second 2D map having an elliptic point. In a later subsection, we also discuss examples of the hyperbolic--hyperbolic (HH) type, where the origin is unstable in both 2D maps of (\ref{4DMMmap}). Note that the case EH is entirely analogous to HE due to the symmetrical form of the two 2D maps.

Throughout our study, we use $10^6$ random initial conditions for each of the variables, i.e. $x_0,x_1$ and $y_0,y_1$, within the domain $(0,10^{-6})$ close to the origin. To facilitate the visualization of stickiness phenomena, observe the phase plane picture shown in Fig. \ref{fig0}(b). The ``warm'' colors represent the more dense parts of the plot, where solutions stick around for very long times, whereas ``cold'' colors depict orbits that scatter diffusively in phase space. We also show in Fig. \ref{fig0}(c) projections of the orbits in the $y_n,y_{n+1}$ plane, which rotate around the origin due to our choice of $K_2<1$.

Each of our initial conditions is iterated $2 \cdot 10^5$ times, to achieve reliable statistics. To obtain the results shown in Fig. \ref{fig0}, we have employed appropriate statistical techniques (see e.g. \cite{rossi,shalizi1}) to optimize both the specific class of suitable pdfs and their parameters to obtain the best fit for such large data sets.

Clearly, a crucial role in this study is played by the fixed point at the origin and its stability properties. 
A simple linearization of the equations of our 4DMM map (\ref{4DMMmap}) about $x_n=y_n=0$ shows that the conditions for stability of the central fixed point with respect to deviations in $x_n$ and/or $y_n$ are: 
\begin{equation}\label{stability}
|K_i+\mu /2 | < 1, \,\,\,\,\, i=1,2
\end{equation}

Thus, we identify as EE (doubly elliptic) the case when both conditions $i=1,2$ in eq. (\ref{stability}) hold, EH (elliptic hyperbolic) if the $i=2$ inequality is reversed, HE (hyperbolic elliptic) if the $i=1$ inequality in (\ref{stability}) is inverted, and HH (doubly hyperbolic), when both inequalities in eq. (\ref{stability}) are reversed. Clearly, if the origin is doubly elliptic (EE), it will be surrounded mostly by quasiperiodic orbits and no large scale chaos will be present in its vicinity.  Hence, in what follows, we will study both ``partly'' unstable HE and ``fully'' unstable cases of the HH type. We start with both $K_i$ positive, but will also consider cases with $K_i<0$, for $i=1,2$. 

\subsection{HE Cases of the 4DMM map}

We begin with a hyperbolic elliptic (HE) case of the 4DMM map problem (\ref{4DMMmap}), with the main parameters chosen so that the $x$-map has $K_1>1$, and the $y$-map $0<K_2<1$, i.e hyperbolic in the $x_n$ plane and elliptic in the $y_n$ plane. 

\begin{figure}
\centering
\includegraphics[height=0.18\textheight]{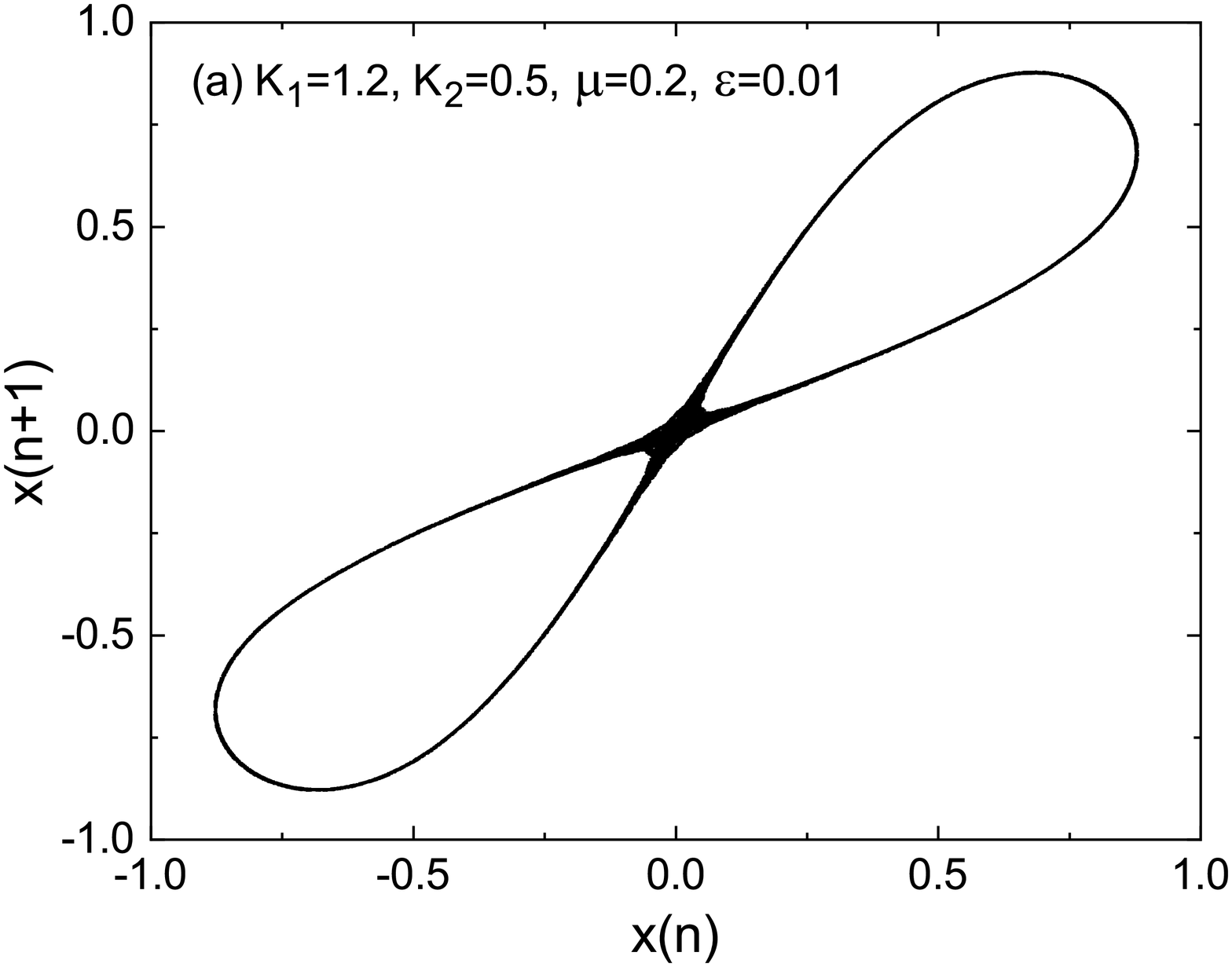}
\includegraphics[height=0.18\textheight]{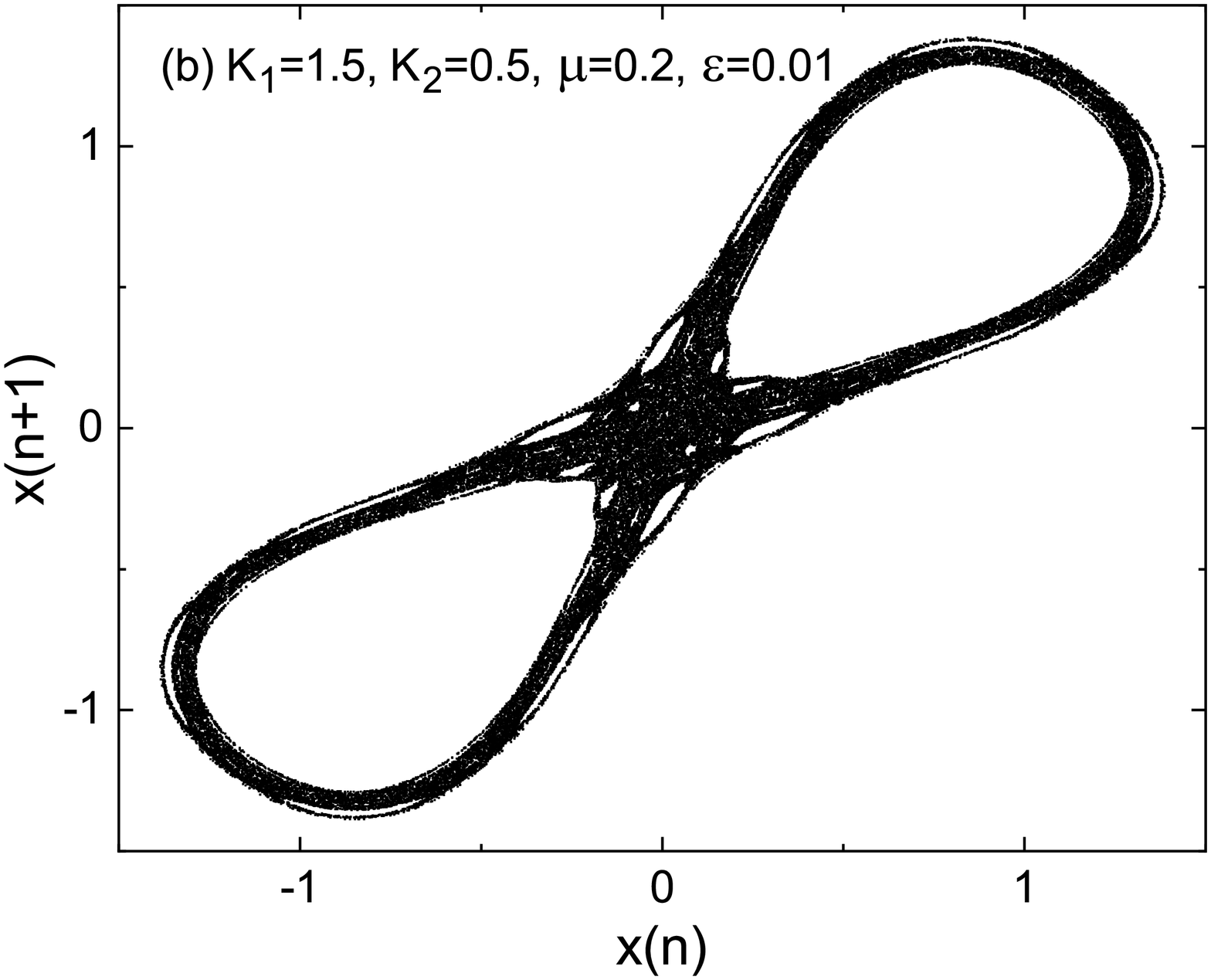}
\includegraphics[height=0.18\textheight]{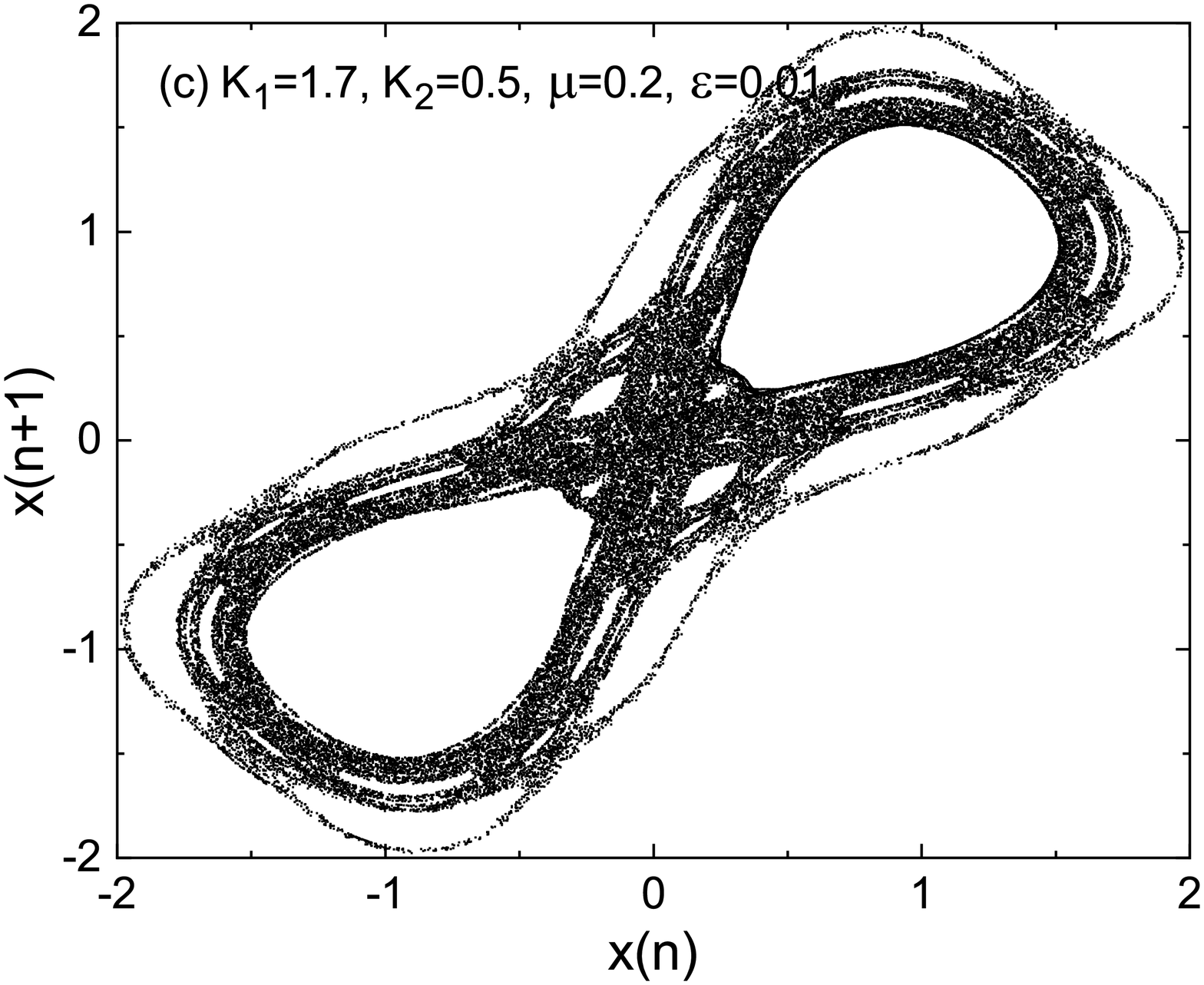}
\includegraphics[height=0.18\textheight]{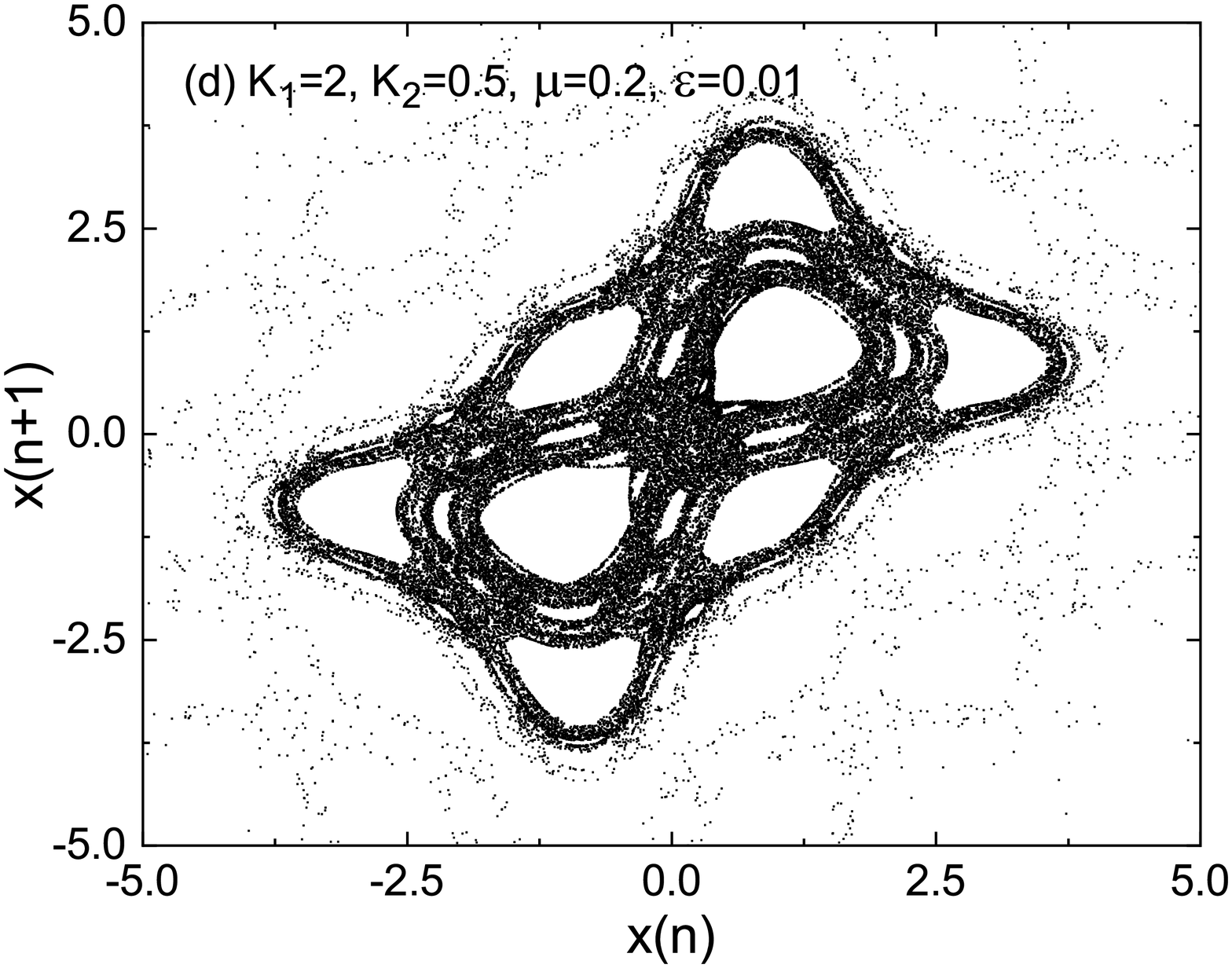}
\caption{2D phase space plots for the $x_{n}$ plane for different $K_1$ values. 
The rest of the parameters, $K_2=0.5, \mu=0.2$ and $\epsilon=0.01$ remain constant for all panels. 
(a) $K_1=1.2$,  (b) $K_1=1.5$, (c) $K_1=1.7$,  and (d) $K_1=2$. The number of iterations is always $2\times10^5$.\label{fig1}}
\end{figure}

Setting $K_2=0.5$ and gradually increasing the value of $K_1$ we observe that the thin `figure--eight' of Fig. \ref{fig1}(a) thickens around the origin as chaos slowly expands, and eventually occupies a wider ``cellular'' domain in phase space shown in Fig. \ref{fig1}(d). 

\begin{figure}
\centering
\resizebox{0.8\columnwidth}{!}{
\includegraphics{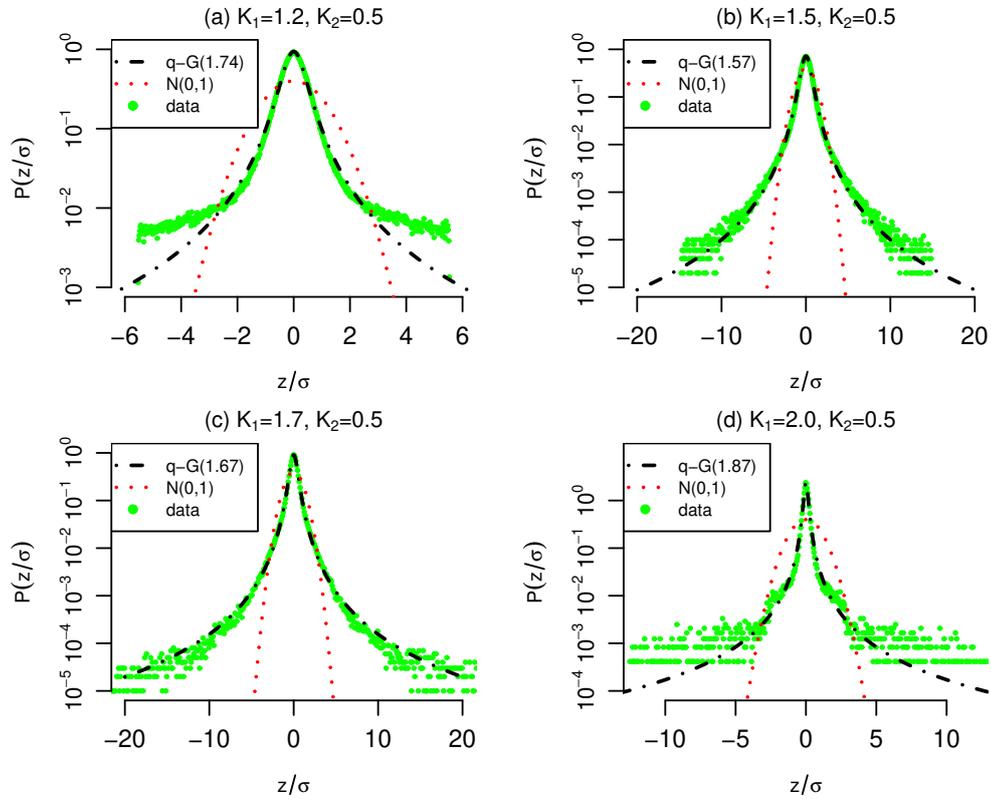}}
\caption{ The pdfs for the sums  $s_N^{(j)}$ corresponding to the chaotic domains shown in Fig. \ref{fig1}(a), (b), (c) and (d) respectively. The black dashed line corresponds to the optimal fitting with the $q$--Gaussian distribution and the red dashed line is the normal distribution.}\label{fig2}
\end{figure}

The pdfs for each of the panels in Fig. \ref{fig1} are depicted in Fig. \ref{fig2}. We observe that as the trajectory winds around a thin figure--eight in Fig. \ref{fig1}(a) in a nearly organized manner, the corresponding distributions of the sums $s_N^{(j)}$ displayed in Fig. \ref{fig2}(a) follow a $q$--Gaussian function for two orders of magnitude, while the tails of the pdf diverge to higher values. The presence of weak chaos, however, for $K_1=1.5,1.7$ in Fig. \ref{fig1}(b) and \ref{fig1}(c) leads to the emergence of optimal $q$--Gaussian distributions in \ref{fig2}(b) and \ref{fig2}(c), which, for $q=1.57, 1.67$, respectively, describe well the numerical data for five orders of magnitude! 

On the other hand, for a higher $K_1=2$ value (see Fig. \ref{fig1}(d)) where the orbits form complex ``cellular'' structures, the $q$-Gaussian distribution that best describes the data in Fig. \ref{fig2}(d) is successful only over two orders of magnitude and corresponds to $q=1.87$. It appears, therefore, that with increasing $K_1$ the value of $q$ increases also.

\begin{figure}
\centering
\includegraphics[scale=0.22]{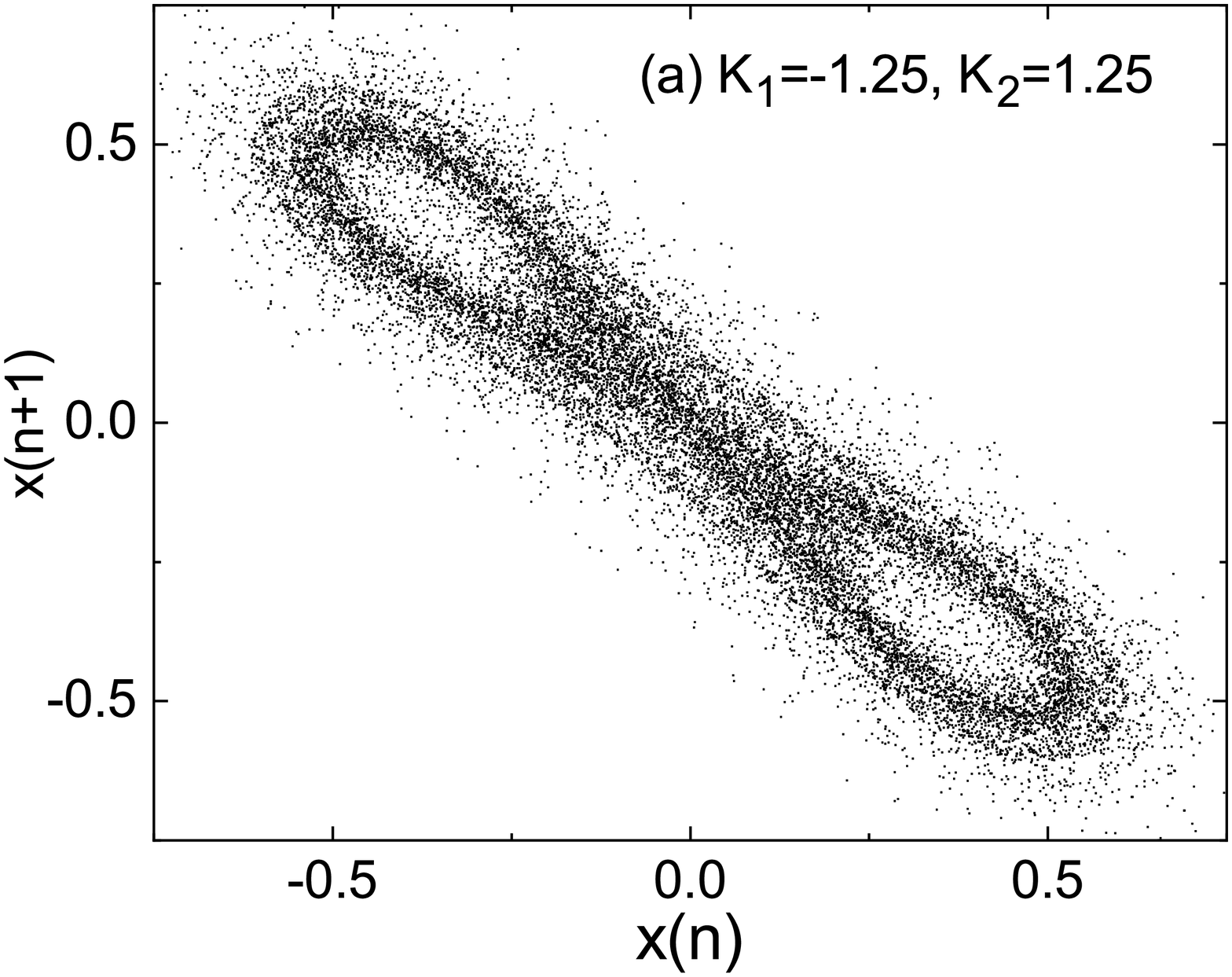}
\includegraphics[scale=0.22]{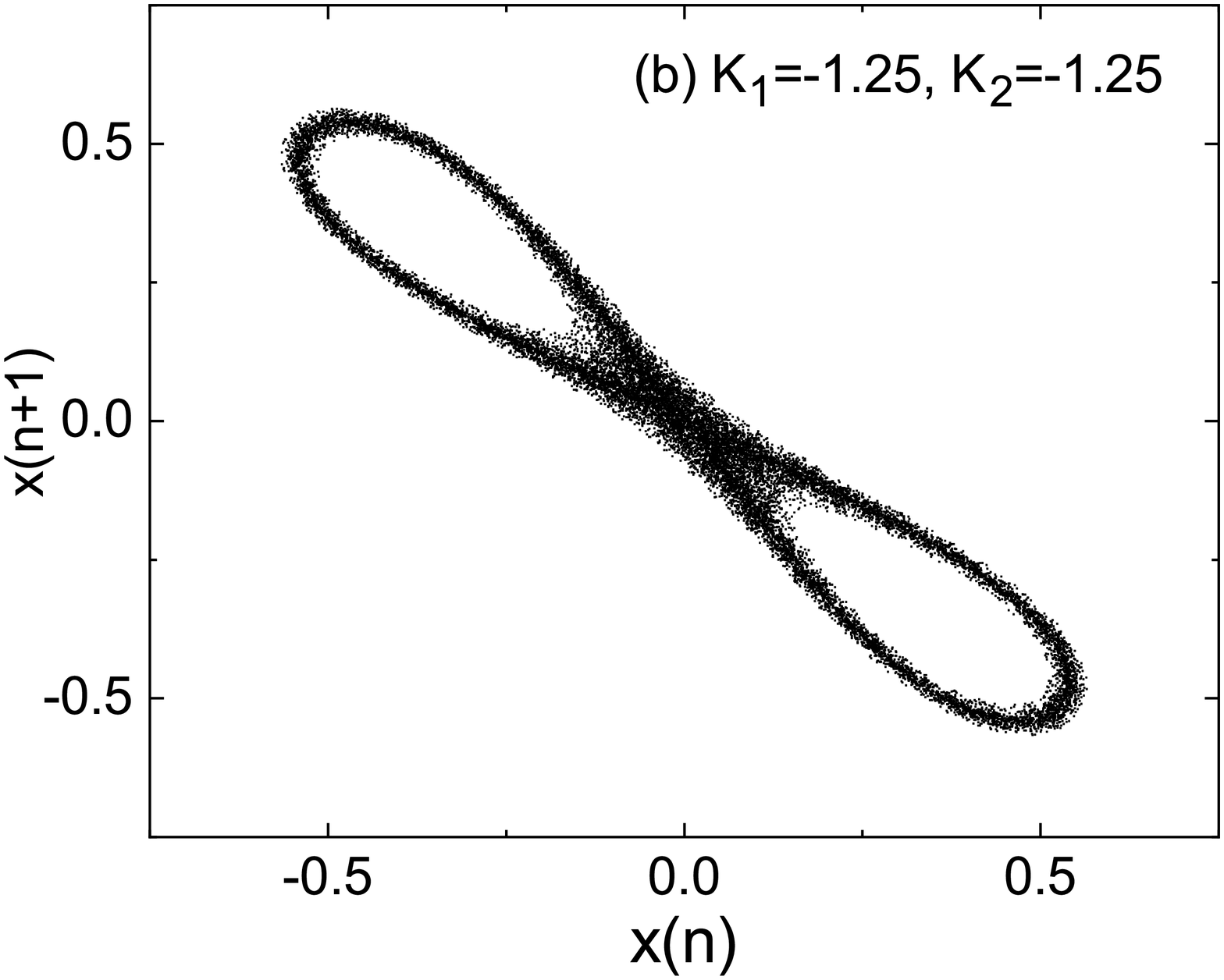}
\includegraphics[scale=0.70]{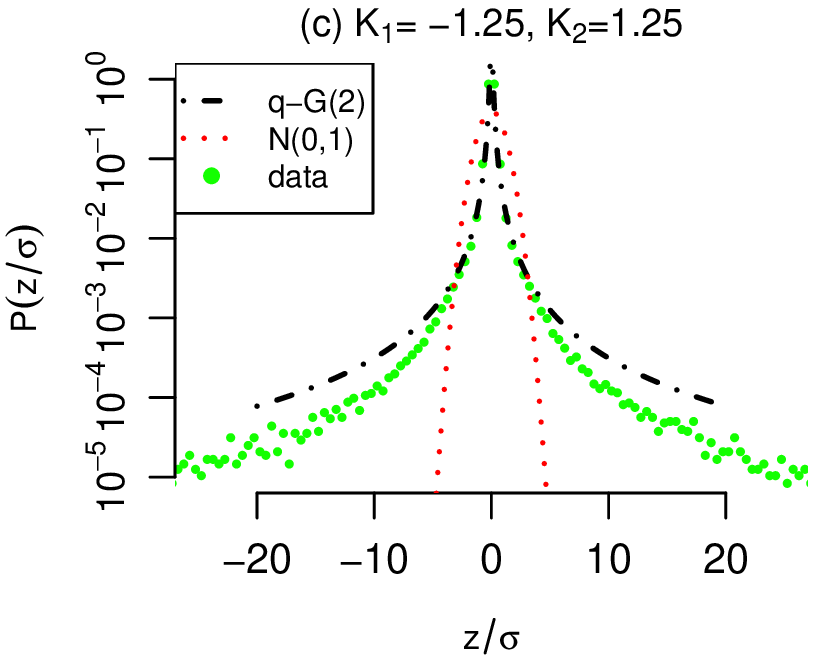}
\includegraphics[scale=0.70]{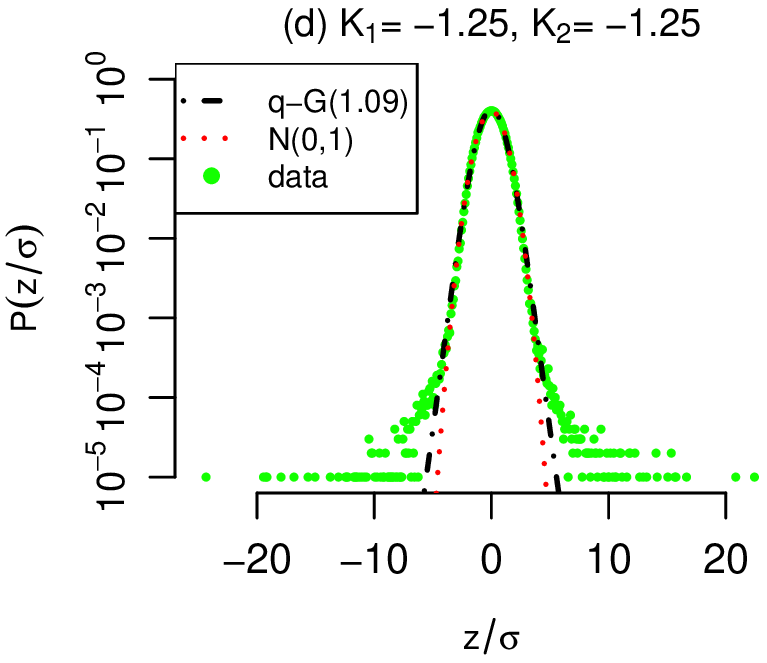}
\caption{Top row: Phase space plots on the $x_{n}, x_{n+1}$ plane for (a) $K_1=-1.25, K2=1.25$, (b) $K_1=K2=-1.25$. Bottom row: (c) and (d) present the pdf plots corresponding to (a) and (b) respectively.\label{fig4}}
\end{figure}

\begin{figure}
\centering
\includegraphics[scale=0.22]{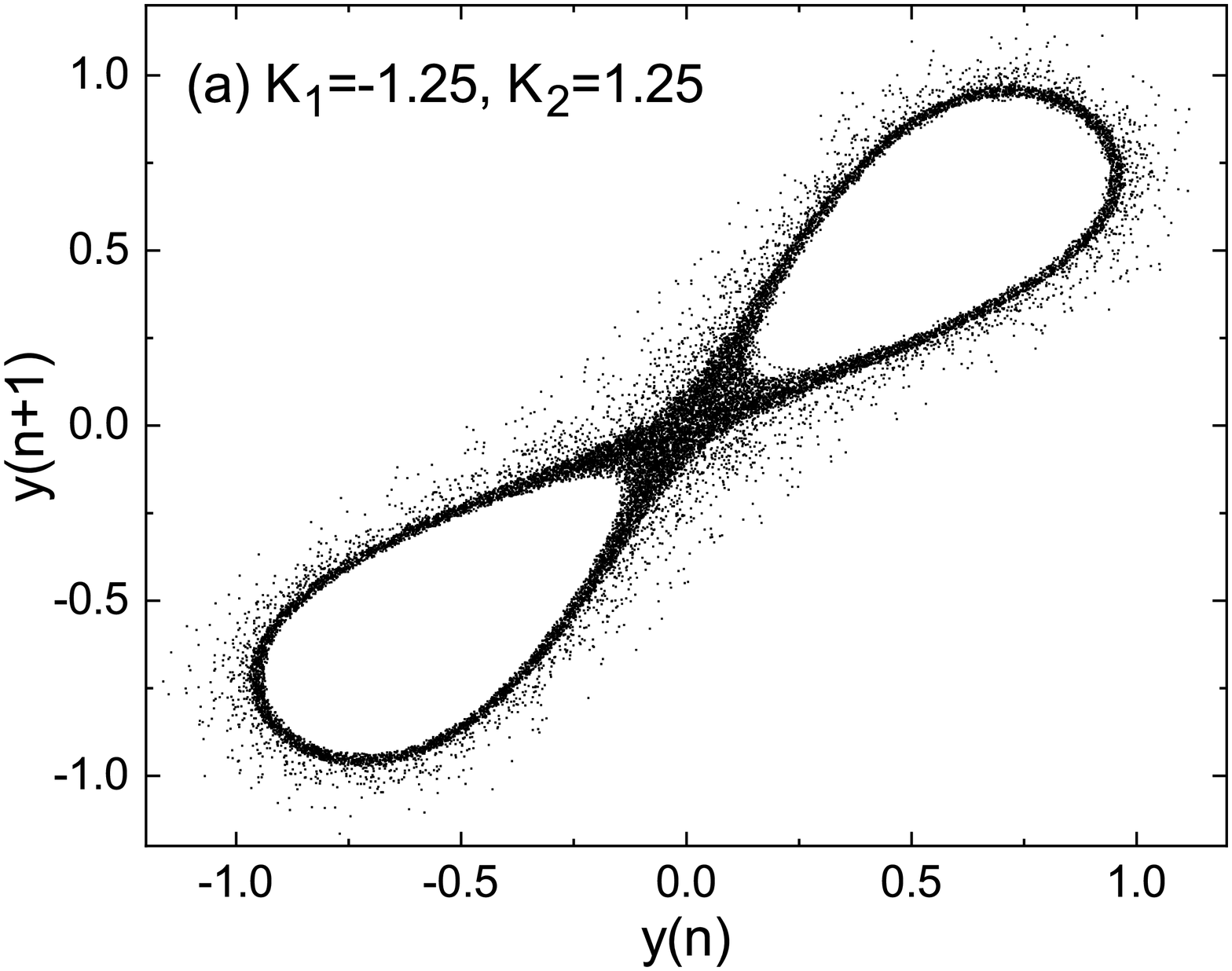}
\includegraphics[scale=0.22]{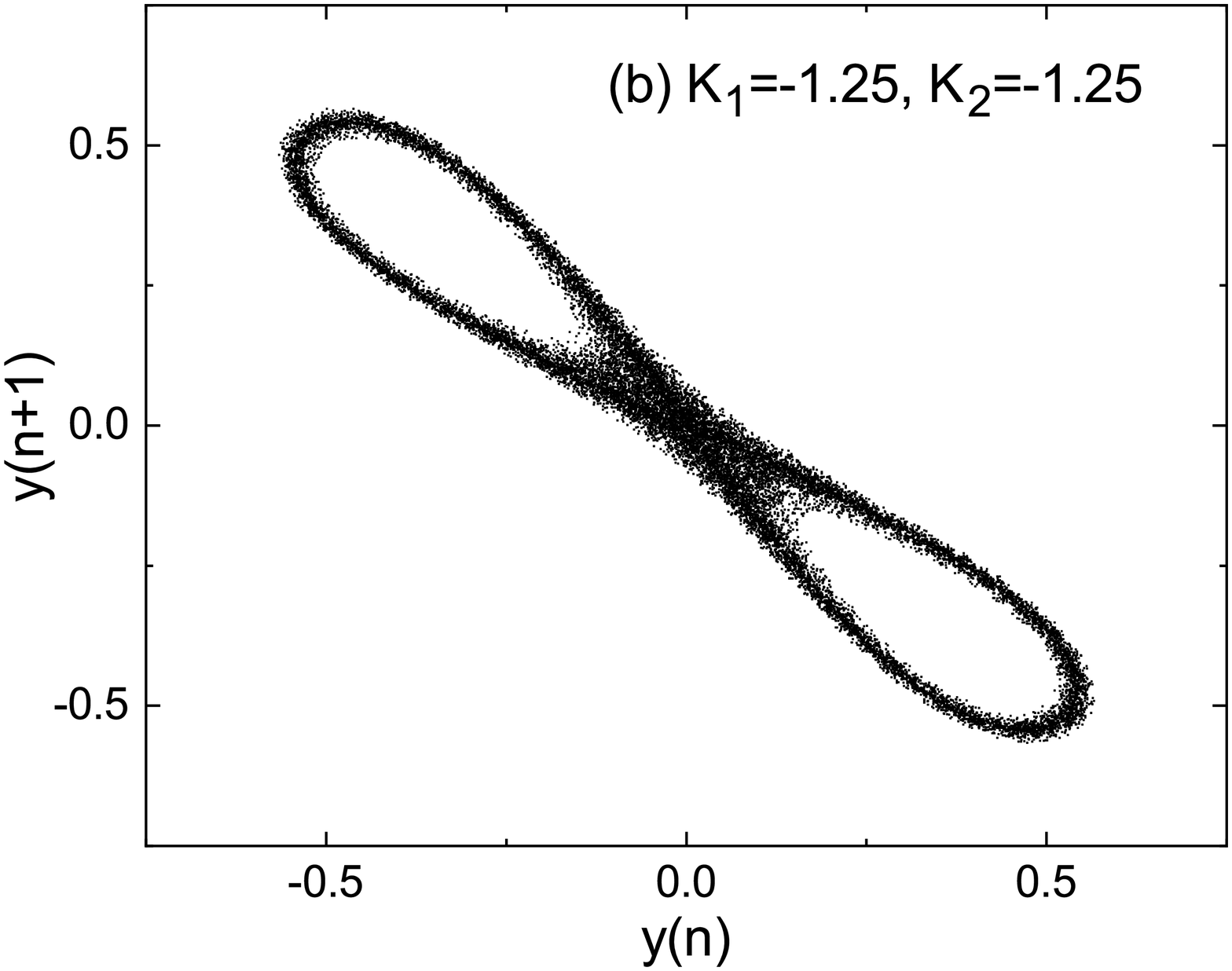}
\includegraphics[scale=0.70]{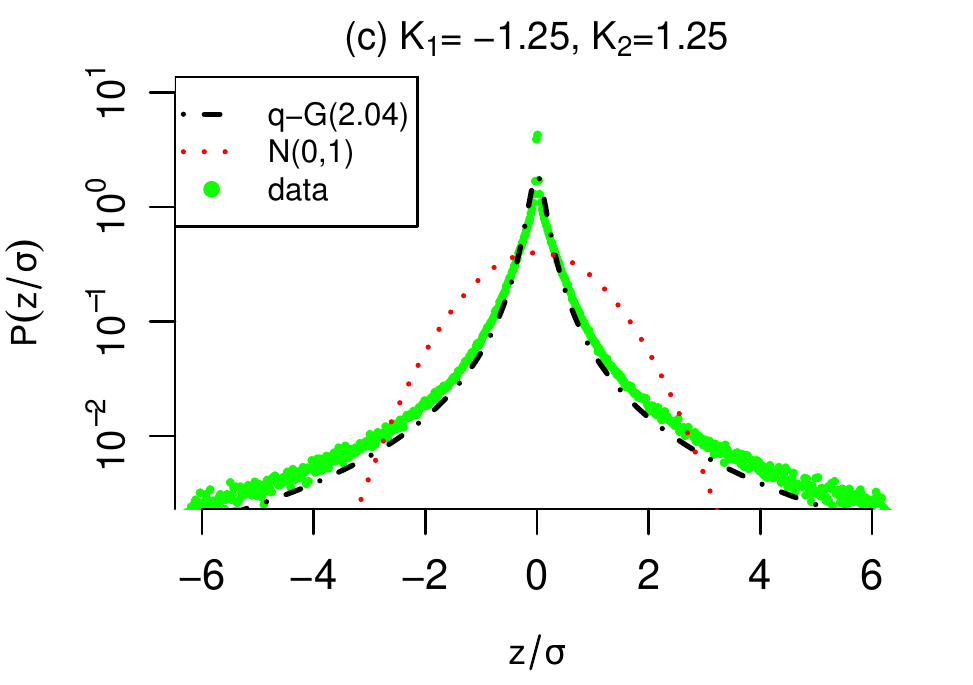}
\includegraphics[scale=0.70]{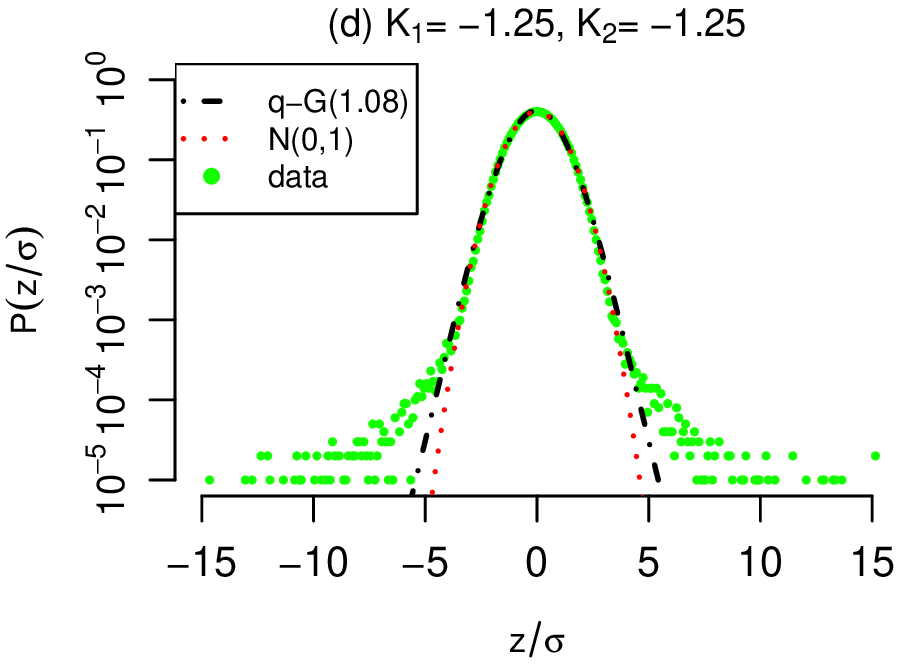}
\caption{(a) Top row: Phase space plots on the $y_{n},y_{n+1}$ plane for (a) $K_1=-1.25, K2=1.25$ and (b) $K_1=K2=-1.25$. In (c) and (d) respectively we plot the pdfs corresponding to (a) and (b). Note the similarities with Fig. \ref{fig4} above.\label{fig5}}
\end{figure}

\subsection{HH cases of the 4DMM map}
Let us now describe some results obtained when the origin of the map is ``fully unstable'', i.e. a double saddle point, which we call hyperbolic-hyperbolic (HH). To this end, we will take values of $K_1$ and $K_2$ that violate the condition (\ref{stability}) and are either positive and negative or both negative as follows: 

1) $K_1=-1.25, K_2= 1.25$: The dynamics is close to weak chaos, as the phase space plot in Fig. \ref{fig4} (a) shows, since its pdf in Fig. \ref{fig4}(c) is close to a $q$-Gaussian for three orders of magnitude with $q=2.97$.

2) With $K_1=-1.25, K_2=-1.25$: The phase space plot in Fig. \ref{fig4} (b) corresponds to what we call ``strong'' chaos, since its pdf, plotted in Fig. \ref{fig4} (d) is very close to a Gaussian with $q=1.09$. 

Observing Fig. \ref{fig4} more closely, we suggest that the  statistical results may be explained as follows: In the first column, where the orbits form a more ``sparse'' pattern in Fig. \ref{fig4}(a), the associated $q$-Gaussian implies weak chaos, while in the second column, a more uniformly filled pattern in Fig. \ref{fig4}(b) is characterized by a true Gaussian representing strong chaos.

Let us also compare, for these HH cases, the above results, with those corresponding to the $y_n,y_{n+1}$ data as plotted in Fig. \ref{fig5}. Clearly, due to the  $x-y$ symmetry of the map, there are strong similarities between Fig. \ref{fig4} and \ref{fig5}, validating the conclusions of weak chaos on the left column and strong chaos on the right column of the two figures.

\subsection{Close to the instability transition}
We also examined a case close to the transition of instability for one of the maps. In particular, as shown in Fig. \ref{fig6} below, we set $K_1=1$ and plot for $K_2=0.9,1.3,1.5$ in Fig. \ref{fig6}(a,c,e) the $x_n,x_{n+1}$ projections of the orbits, while in Fig. \ref{fig6}(b,d,f) we present the corresponding statistical analysis. Clearly the pdfs in this case are very well described by a $q$-Gaussian with $q$ increasing from $1.5$ to $1.94$ and $2.04$, close to the value $q=2$, which is the case of the Cauchy distribution.

\begin{figure}
	\centering
	\includegraphics[scale=0.2]{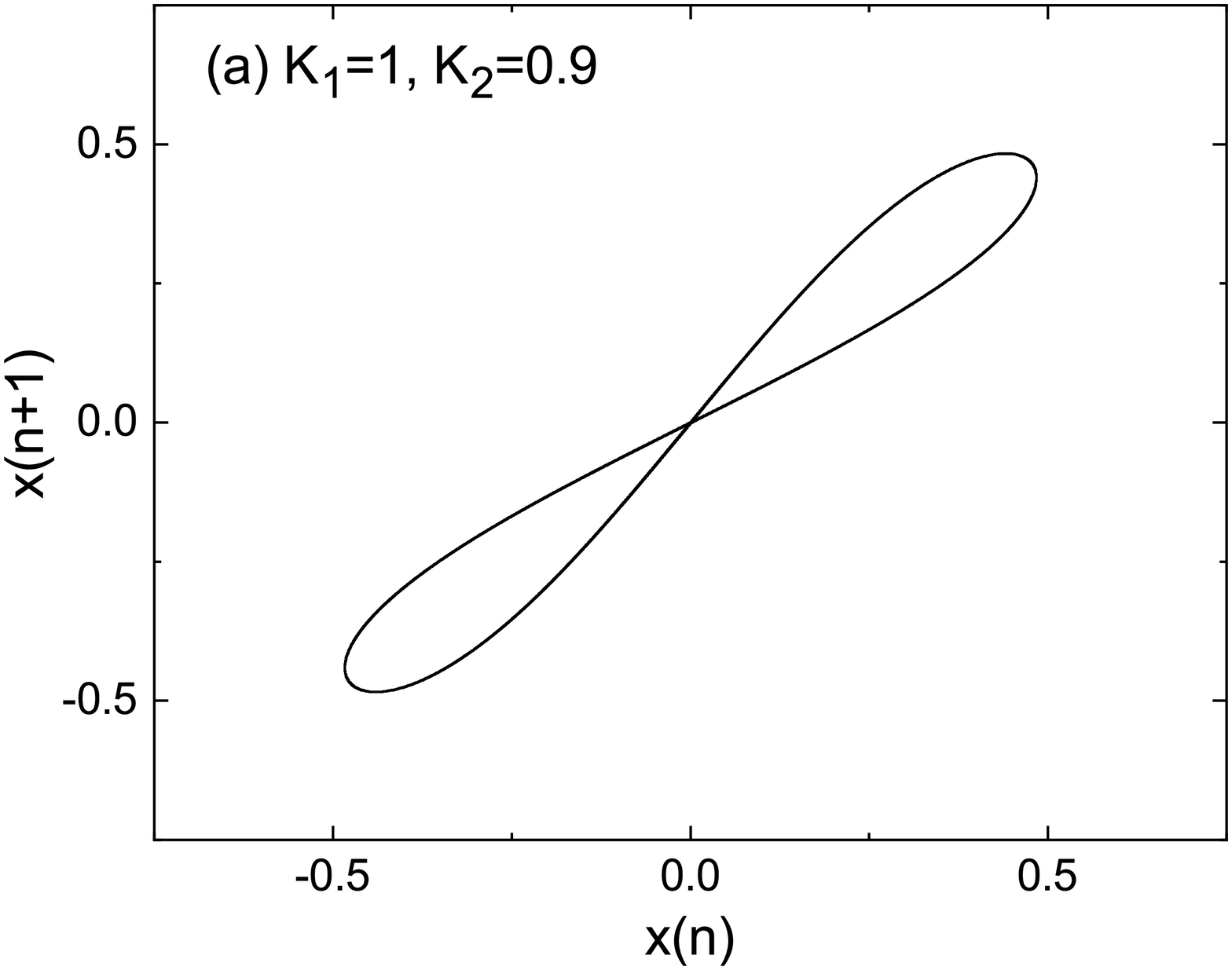}
	\includegraphics[scale=0.68]{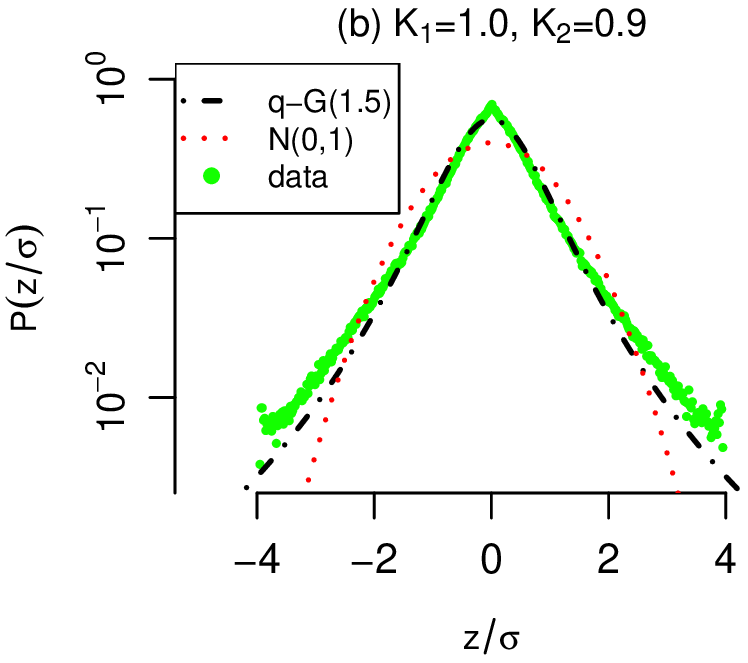}
	\includegraphics[scale=0.2]{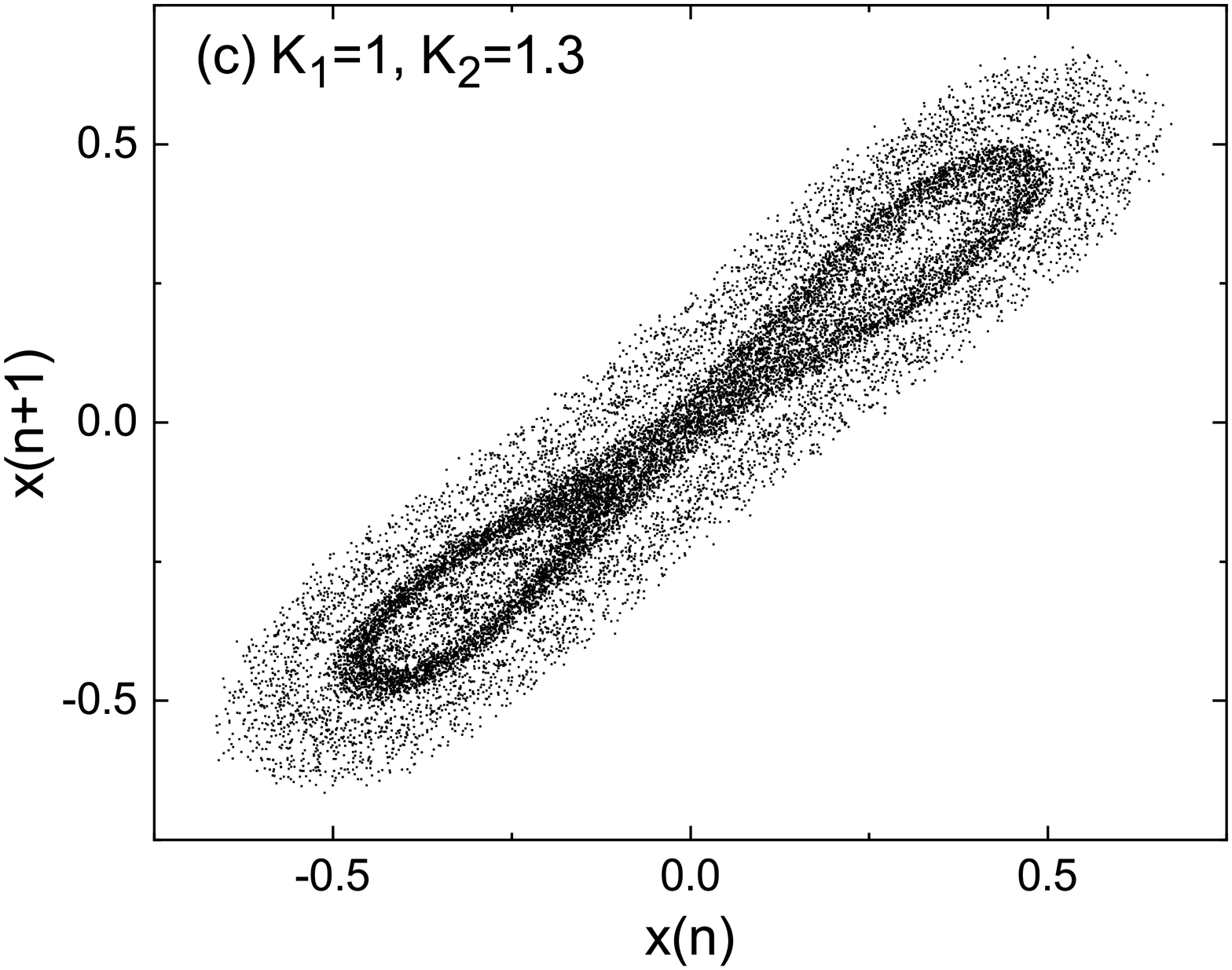}
	\includegraphics[scale=0.68]{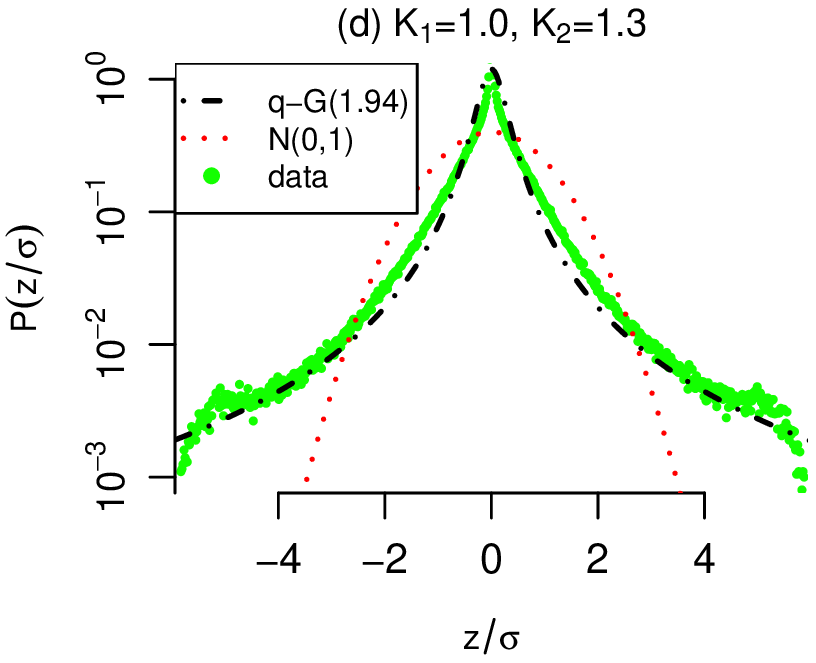}
	\includegraphics[scale=0.2]{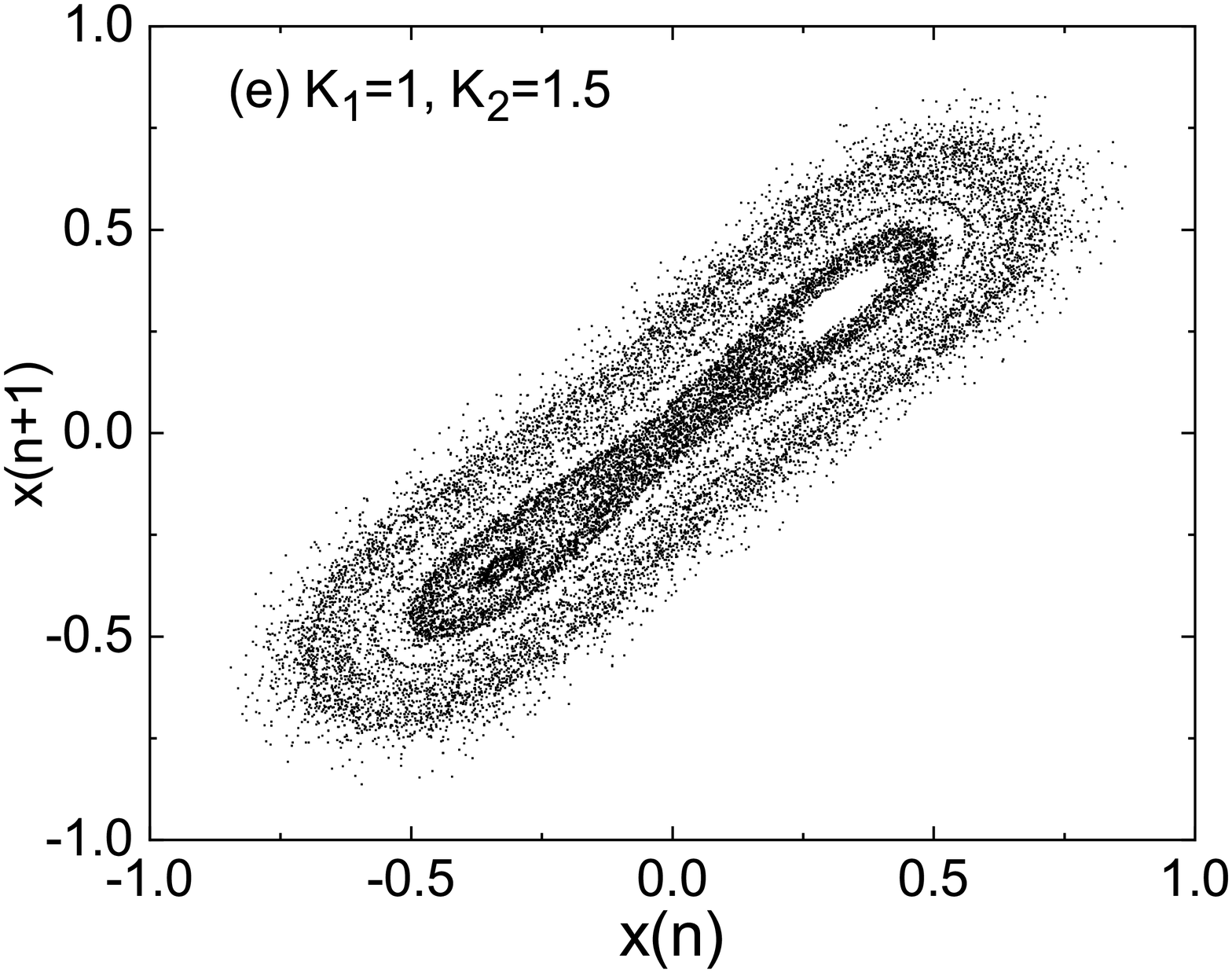}
	\includegraphics[scale=0.68]{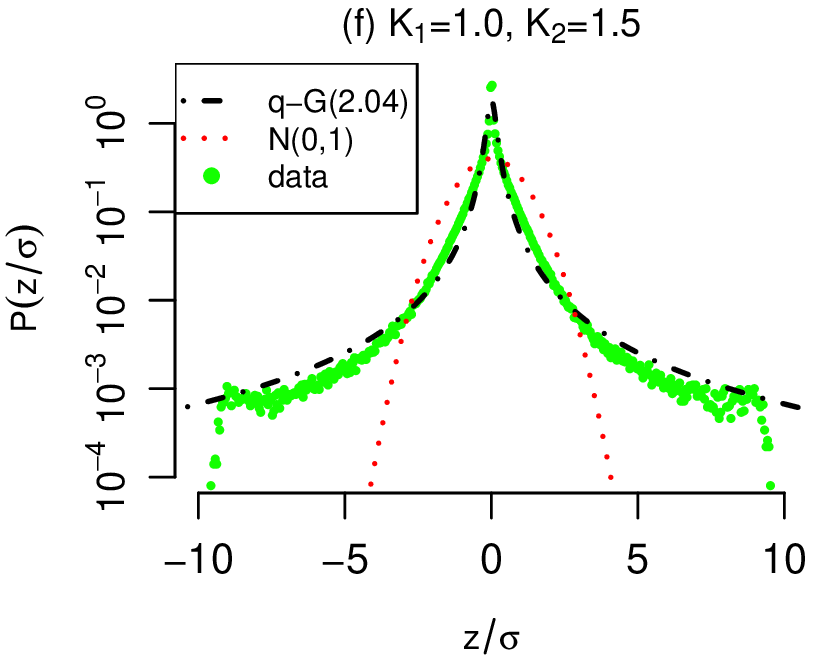}
	\caption{Close to the instability transition: Here we set $K_1=1$ and present in (a)-(c)-(e) the phase space plots for $x_n$ and in (b)-(d)-(f) the corresponding pdf plots. The first row corresponds to $K_2=0.9$, the second row to $K_2=1.3$ and the third row to $K_2=1.5$. \label{fig6}} 	
\end{figure}

%%%%%%%%%%%%%%%%%%%%%%%%%%%%%%%%%%%%%%%%%%%%%%%%%%%%%%%%%%%%%%%%%%%%%%%%%%%%%%%%%%%%

\section{Conclusions\label{Section4}}

The stickiness of orbits observed in the vicinity of unstable periodic orbits of higher dimensional symplectic maps, or Hamiltonian systems of more than $2$ degrees of freedom, is clearly a complex phenomenon. It has been termed ``weak chaos'' in the literature mainly because its statistical analysis reveals that it is associated with $q$-Gaussian probability distributions, as opposed to the simple Gaussians one finds when studying uniformly spread stochasticity associated with Boltzmann Gibbs statistics. This is because the motion in weakly chaotic situations is correlated over long ranges, while in strongly chaotic regions the correlations are short ranged.

In this paper, we attempted to study this phenomenon, for the first time, in a 4-D symplectic map, serving as a paradigm for Hamiltonian systems of 3 degrees of freedom. Our results suggest that ``weak chaos'' arises typically near unstable fixed points of $2N$-dimensional maps and may very well be present also near unstable periodic orbits in higher dimensional settings.

In most examples we considered, chaos tends to form ``organized'' patterns in phase space, while the pdfs describing their statistics attain $1<q<2$ values suggesting the presence of strong correlations in the dynamics. However, we have also observed cases where chaos spreads more uniformly in phase space and $q$ tends to approach the value $q=1$ yielding purely Gaussian distributions.

We also observed that as the main nonlinear parameters of the model $K_i,i=1,2$ increase, the values of the index $q$ of the distributions also grow. However, the genericity of these results remains open and needs to be studied further in more general classes of 4-D symplectic maps.

Clearly, every high--dimensional conservative dynamical system will have its own particular features determining the nature of chaos present near its unstable periodic orbits. We believe, however, that the results presented in this paper suggest that weak chaos is generic and may have important implications regarding the dynamics of higher dimensional conservative systems of physical significance.

\section*{Acknowledgements}
We happily dedicate the present manuscript to Professor Thanassis Fokas on the occasion of his 70th birthday and wish him many more years of pioneering work in all sciences. We thank the referees for their interesting suggestions and remarks. T. Bountis acknowledges the hospitality of the NCSR "Demokritos" and many discussions with colleagues at the Institute of Nanoscience and Nanotechnology. This work was supported by the Russian Science Foundation (project No. 21-71-30011), https://rscf.ru/en/project/21-71-30011/. T. Bountis acknowledges financial support for Sections 1, 3.2, 3.3 and 4 of the paper.

\bibliographystyle{plain}
\bibliography{MM4dmaps}

\end{document}